\documentstyle[aps,pra,epsf,preprint]{revtex}
\begin{document}
\draft
\title{Mixedness and teleportation}
\author{S. Bose  and  V. Vedral}
\address{Centre for Quantum Computation, Clarendon Laboratory,
	University of Oxford,
	Parks Road,
	Oxford OX1 3PU, England}

\maketitle
\begin{abstract}
We show that on exceeding a certain degree of mixedness (as quantified
by the von Neumann entropy), entangled
states become useless for teleporatation. By increasing the
dimension of the entangled
systems, this entropy threshold can be made arbitrarily close to 
maximal. This entropy is found to exceed the entropy 
threshold sufficient to ensure the failure of dense coding.
\end{abstract}


 Shared bipartite entanglement has found a host of interesting
applications in quantum communications \cite{Ek,wies,ben}. It is natural to
expect that the efficiency of these applications would go down
with the decrease of shared entanglement. However, apart from
the degree of entanglement of a shared state, there is another physical factor, namely the
mixedness of the state, which causes deterioration of the efficiency
of the applications. Though for given classes of states (such as
the Werner states \cite{wer}), the entanglement
of the state may decrease with the mixedness of the state, the two are
not necessarily related concepts. For example, a mixed state
can have more entanglement than a completely pure (zero
mixedness) disentangled state. Thus we are interested in how the
mixedness of a given state, taken as an independent
physical criterion, affects the efficiency of the entanglement
applications. In particular, we will focus on teleportation \cite{ben}.
   
  A good measure of mixedness of a state
$\rho$ is its von Neumann entropy \cite{vN}
$S(\rho)=-\mbox{Tr}(\rho \log \rho)$. We will first show that when
the entropy of a given $N\times N$ state exceeds $\log{N}+(1-\frac{1}{N})\log(N+1)$, the state becomes useless
for teleportation. To this end we will first need to prove a short theorem.
For this theorem we need a quantity called the singlet fraction
introduced by the Horodeckis \cite{hor1}. The singlet fraction
$F(\rho)$ of a $N\times N$ state $\rho$ is defined as $\mbox{max}\langle \Psi| \rho  |\Psi\rangle$, where the maximum is taken over all the $N\times N$
maximally entangled states. We now proceed to our theorem.
 
  {\bf Theorem:}{\em If the entropy $S(\rho)$ of a state $\rho$ of a
$N\times N$ system exceeds $\log{N}+(1-\frac{1}{N})\log(N+1)$, then the  singlet fraction $F(\rho)< \frac{1}{N}$}. \\
{\bf Proof:} Let, for a certain state $\rho$, $F(\rho)\geq \frac{1}{N}$.
This means that there exists, at least one $N\times N$ maximally entangled state
$|\Psi_{\mbox {\scriptsize Max}}\rangle$, for which $\langle \Psi_{\mbox {\scriptsize Max}}|\rho |\Psi_{\mbox {\scriptsize Max}}\rangle \geq \frac{1}{N}$. Let  us write the state $\rho$ as
\begin{equation}
\rho=\sum_{i=1,j=1}^{N^2} c_{ij}|i\rangle \langle j|,
\end{equation}
where $\{|i\rangle\}$ is a basis formed from $|\Psi_{\mbox {\scriptsize Max}}\rangle$  and $N^2-1$ other maximally entangled states. From the definition
of singlet fraction it follows that the largest of the elements $c_{ii}$
(say this is $c_{11}$) has a value greater than or equal to $\frac{1}{N}$.
 Now, we
know that the von Neumann entropy $S(\rho)$ of the state $\rho$ is always 
less than or equal to its Shannon entropy in any particular basis. This implies
\begin{equation}
S(\rho) \leq -\sum_{i=1}^{N^2} c_{ii}\log c_{ii}.
\label{e1}
\end{equation}
Subject to the constraint $c_{11} \geq \frac{1}{N}$, the expression
$-\sum_{i=1}^{N^2} c_{ii}\log c_{ii}$ attains its highest value when
$c_{11}=\frac{1}{N}$ and the rest $N^2-1$ elements $c_{ii}$ are all equal.
Thus
\begin{eqnarray}
-\sum_{i=1}^{N^2} c_{ii}\log c_{ii} \leq &-&\frac{1}{N}\log \frac{1}{N}
\nonumber \\ &-&(1-\frac{1}{N})\log \{ \frac{1}{N^2-1}(1-\frac{1}{N})\}  
\nonumber \\ &=& \log{N}+(1-\frac{1}{N})\log(N+1).
\label{e2}
\end{eqnarray}
From Eqs.(\ref{e1}) and (\ref{e2}) it follows that
\begin{equation}
S(\rho) \leq \log{N}+(1-\frac{1}{N})\log(N+1).
\end{equation}
Thus we have
\begin{equation}
F(\rho)\geq \frac{1}{N} \Longrightarrow S(\rho) \leq \log{N}+(1-\frac{1}{N})\log(N+1).
\label{e3}
\end{equation}
The implication in the above equation is equivalent to
\begin{equation} 
S(\rho) > \log{N}+(1-\frac{1}{N})\log(N+1)
\Longrightarrow F(\rho) < \frac{1}{N}.
\end{equation}

  In Ref.\cite{hor1} the Horodeckis have shown that singlet
fraction $F (\rho) < \frac{1}{N}$ implies that one cannot do teleportation
with $\rho$ with better than classical fidelity. Thus when the
entropy of a state exceeds $\log{N}+(1-\frac{1}{N})\log(N+1)$, then by
virtue of the theorem proved above, the state becomes useless for
teleportation. Here, the phrase "useless for teleportation" means "useless for
teleportation with better than classical fidelity". Note that this
value of entropy is a minimum threshold. At values of entropy arbtrarily
close to this but less, a state $\rho$ is not forbidden to allow better than classical teleportation. For example, consider the generalized Werner
state \cite{mio} $W_N(\epsilon)=\epsilon |\Psi_N \rangle \langle \Psi_N| +(1-\epsilon)
\rho_{\scriptsize{\mbox{M}}}$ of $N \times N$ dimensions where $\rho_{\scriptsize{\mbox{M}}}$ is
the corresponding maximally mixed state. When $\epsilon$ is infinitisimally
greater than $\frac{1}{N}$ (which automatically ensures that the singlet fraction is $>\frac{1}{N}$)
the state will allow teleportation better
than classical, but its entropy will only be slightly below $\log{N}+(1-\frac{1}{N})\log(N+1)$. 
    
  An interesting consequence of our
result is the fact that as the dimension $N$ of the systems is
increased, the entropy threshold becomes closer and closer to the
maximal possible entropy of the state. In fact as $N \rightarrow \infty$,
we have $\log{N}+(1-\frac{1}{N})\log(N+1) \rightarrow 2 \log N$. Thus
for systems of very large dimensions, {\em even an entropy extremely
close to the maximal entropy is not sufficient to ensure the failure
of teleportation}.

  It is now interesting to compare the entropy sufficient to ensure the
failure of teleportation with the entropy sufficient to ensure the
failure of another application, namely, dense coding \cite{wies}. 
Dense coding with mixed states have been studied before
\cite{us,ben2}, but here our target is to identify a degree
of mixedness above which dense coding is bound to fail. Here again,
failure of dense coding will mean its capacity being less
than or equal to the classical communication capacity of $\log N$
bits per qu-N-bit. An upper bound to the capacity for dense coding with
mixed signal states $W_i$ occurring with probabilities $p_i$
is given by the Holevo bound \cite{khol1} $H=S(\sum p_i W_i)-\sum p_i S(W_i)$.
The first expression $S(\sum p_i W_i)$ can attain at most a value
of $2 \log N$. Thus when the entropy $S(W_i)$ of a signal state
exceeds $\log N$ we have $H \leq \log N$. Therefore an entangled state
$\rho$ will fail to be useful for dense coding when $S(\rho) > \log N$.
This is also a minimum threshold. For example, for the state $W_N(\epsilon)$, we
have $H=2 \log N -S(W_N(\epsilon))$ for standard Bennett and Wiesner
scheme of dense
coding and this can exceed $\log N$ for $S(W_N(\epsilon))$ slightly less
than $\log N$.  This threshold of $\log N$ is evidently much smaller
than the threshold $\log{N}+(1-\frac{1}{N})\log(N+1)$ sufficient to
ensure the failure of teleportation. 

      In this paper we have shown that there is a degree
of mixedness after which a state becomes useless for teleportation.
We have quantified this mixedness with the von Neumann entropy, but
we could as well use the linear entropy $S_L=1-\mbox{Tr}\rho^2$. In that case
the threshold for failure of teleportation will be $1-\frac{2}{N(N+1)}$.
The fact that on increasing the mixedness of a state, dense coding
fails before teleportation indicates that teleportation is "more
robust" to external noise. Of course, our entropic criterion is only a {\em sufficient
condition} for the failure of teleportation. However, entropic criteria can never be necessary for the failure
of any entanglement application because they fail even for
 pure disentangled states. 
It would be easier
to calculate the entropy of a state than to calculate its singlet
fraction as no maximization is involved in the former calculation.
Hence mathematically, our entropic criterion ($S > \log{N}+(1-\frac{1}{N})\log(N+1)$) is more convinient than
the corresponding singlet fraction condition ($F<\frac{1}{N}$).
How about the realtion between mixedness and entanglement itself?
We know that for a Bell diagonal state $\rho$ with only two non-zero
eigenvalues, the distillable entanglement \cite{pur} is equal to
$1-S(\rho)$ \cite{ved}. Such a state would not be
distillable if $S(\rho) \geq 1$.   
Is there such an entropy threshold sufficient to ensure the failure
of entanglement distillation for an arbitrary $N \times N$ state? 
We leave that as an interesting
open question.


\begin{references}
%
\bibitem{Ek} A. K. Ekert, Phys. Rev. Lett. {\bf 67}, 661 (1991).
%
%
\bibitem{wies} C. H. Bennett and S. J. Wiesner,  Phys. Rev. Lett. {\bf 69} , 2881 (1992).
%
\bibitem{ben} C. H. Bennett, G. Brassard, C. Crepeau, R. Jozsa, A. Peres and W. K. Wooters, Phys. Rev. Lett. {\bf 70}, 1895 (1993).
%
\bibitem{wer}
R. F. Werner, Phys. Rev. A {\bf 40}, 4277 (1989).
%
\bibitem{vN}
J. von Neumann, {\em Mathematical Foundations of Quantum Mechanics},
Princeton University Press, Princeton (1955) (Eng. Trans.
by R. T. Beyer).
%
\bibitem{hor1}
M. Horodecki, P. Horodecki and R. Horodecki, Phys. Rev. A {\bf 60}, 1888 (1999).
%
\bibitem{mio} M. Murao, M. B. Plenio, S. Popescu, V. Vedral and P. L. Knight,
Phys. Rev. A {\bf 57}, R4075 (1998).
%
\bibitem{us}
S. Bose, M. B. Plenio and V. Vedral, LANL eprint: 
quant-ph/9810025 (to appear in J. Mod. Opt).
%
\bibitem{ben2}
C. H. Bennett, P. W. Shor, J. A. Smolin and A. V. Thapliyal,
Phys. Rev. Lett. {\bf 83}, 3081 (1999).
%
\bibitem{khol1}
A. S. Kholevo, Probl. Peredachi Inf {\bf 9}, 177 (1973) [A. S. Kholevo, Problems of Information Transmission, {\bf 9}, 177 (1973)].
%
\bibitem{pur}
C. H. Bennett, D. P. DiVincenzo, J. A. Smolin  and W. K. Wootters,
Phys. Rev. A {\bf 54}, 3824 (1996).
%
\bibitem{ved}
V. Vedral and M. B. Plenio,  Phys. Rev. A {\bf 57}, 1619 (1998);
E. M. Rains, Phys. Rev. A {\bf 60}, 179 (1999).
\end{references}
\end{document}